\PassOptionsToPackage{prologue,table}{xcolor}
\documentclass[sigconf]{acmart}
\AtBeginDocument{%
  }

\setcopyright{acmlicensed}
\copyrightyear{2025}
\acmYear{2025}
\acmDOI{XXXXXXX.XXXXXXX}
\acmConference[KDD '25]{Agentic \& GenAI Evaluation}{Aug 4,
  2025}{Toronto, ON}
\acmISBN{978-1-4503-XXXX-X/2018/06}

\usepackage{todonotes}
\usepackage{enumitem}
\usepackage[linesnumbered,ruled,vlined]{algorithm2e}
\newlist{rqlist}{enumerate}{1}
\setlist[rqlist,1]{label={RQ\arabic*}}
\usepackage{booktabs}
\usepackage{colortbl}
\usepackage{graphicx}
\usepackage{subcaption}
\usepackage{float}
\usepackage{multirow}
\usepackage{graphicx}
\usepackage{tabularx}
\usepackage{hyperref}
\newcommand{\forepbench}{\textsc{FoRepBench}}
\newcommand{\bootstrapgen}{\textsc{Bootstrap Generator}}
\newcommand{\llmvalidator}{\textsc{LLM Validator}}

\begin{document}

\title{Benchmark Dataset Generation and Evaluation for Excel Formula Repair with LLMs}
\author{Ananya Singha}
\authornote{Both authors contributed equally to this research.}
\affiliation{
\institution{Microsoft}
\country{United States}
}
\email{ananyasingha@microsoft.com}

\author{Harshita Sahijwani}
\authornotemark[1] % Refers to the first authornote above
\affiliation{
\institution{Microsoft}
\country{United States}
}
% \affiliation{%
\email{hasahijwani@microsoft.com}

\author{Walt Williams}
\affiliation{
\institution{Microsoft}
\country{United States}
}
\email{walwilliams@microsoft.com}
% \affiliation{%

\author{Emmanuel Aboah Boateng}
\affiliation{
\institution{Microsoft}
\country{United States}
}
\email{emmanuelab@microsoft.com}
% \affiliation{%

\author{Nick Hausman}
\affiliation{
\institution{Microsoft}
\country{United States}
}
% \affiliation{%
\email{nhausman@microsoft.com}

\author{Miguel Di Luca}
\affiliation{
\institution{Microsoft}
\country{United States}
}
% \affiliation{%
\email{migueldiluca@microsoft.com}

\author{Keegan Choudhury}
\affiliation{
\institution{Microsoft}
\country{United States}
}
\email{kechoudhury@microsoft.com}

\author{Chaya Binet}
\affiliation{
\institution{Microsoft}
\country{United States}
}
\email{Chaya.Leiser@microsoft.com}

\author{Vu Le}
\affiliation{
\institution{Microsoft}
\country{United States}
}
\email{levu@microsoft.com}

\author{Tianwei Chen}
\affiliation{
\institution{Microsoft}
\country{United States}
}
\email{tianweichen@microsoft.com}

\author{Oryan Rokeah Chen}
\affiliation{
\institution{Microsoft}
\country{United States}
}
\email{orokeahchen@microsoft.com}

\author{Sulaiman Vesal}
\affiliation{
\institution{Microsoft}
\country{United States}
}
% \affiliation{%
\email{svesal@microsoft.com}

\author{Sadid Hasan}
\affiliation{
\institution{Microsoft}
\country{United States}
}
\email{sadidhasan@microsoft.com}
% \affiliation{%

\renewcommand{\shortauthors}{}

%% article
% \begin{abstract}
% \todo[inline]{Need to add author names as the submission is single blind. Todo: add emails and affiliations if space permits}
% Excel formulas can be complex and difficult to work with, particularly for novice users. Foundation models like GPT-4o provide useful assistance for this task by explaining errors within the formula. However, the automatic correction of runtime errors - those arising from a user's logical mistakes or misinterpretations of Excel functions - remains an open challenge.
% A key barrier in building models for these scenarios is the lack of high-quality datasets for training and evaluation.
% This paper presents an approach for constructing a benchmark dataset for Excel formula repair.
% % In particular, we introduce two methods for benchmark generation namely {\em Bootstrap Generator} and {\em Adversarial Generator}. Both these methods comprise agents for dataset generation and validation. We evaluate the quality of our generated datasets through manual annotation. 
% \todo[inline]{add key results}
% The proposed methods can be applied to generate evaluation benchmarks for related tasks like debugging low-resource programming languages.
% % We also present initial experiments for automated formula correction using LLMs. 

% \end{abstract}
\begin{abstract}
Excel is a pervasive yet often complex tool, particularly for novice users, where runtime errors arising from logical mistakes or misinterpretations of functions pose a significant challenge. While large language models (LLMs) offer promising assistance by explaining formula errors, the automated correction of these semantic runtime errors remains an open problem. A primary challenge to advancing models for such scenarios is the severe lack of high-quality, comprehensive datasets for training and rigorous evaluation. This paper addresses this gap by introducing a novel approach for constructing a benchmark dataset specifically designed for Excel formula repair. We propose a data generation pipeline, which leverages a small set of curated seed samples from online forums to synthetically expand the dataset. Our pipeline integrates few-shot prompting with LLMs and employs a robust \textit{LLM-as-a-Judge} validation framework, combined with execution-based checks to ensure the correctness and semantic fidelity of the generated data. This process produced a benchmark dataset of 618 high-quality samples, covering common runtime errors. Furthermore, we propose a context-aware baseline technique for Excel formula repair that utilizes LLMs to leverage both the faulty formula, and relevant spreadsheet context.
We evaluate the performance of various LLMs (GPT-4o, GPT-4.1, Phi-3, Mistral) on our newly generated benchmark using execution-based metrics. Our analysis demonstrates the dataset's quality through manual annotation and provides insights into error and function distributions. The proposed generation methodology is highly scalable and can be readily adapted to create evaluation benchmarks for similar code repair tasks in other low-resource programming languages.
\end{abstract}


\maketitle

\section{Introduction}
\label{sec:introduction}

Spreadsheets (e.g., Microsoft Excel, Google Sheets) are among the most widely used end-user programming platforms, with hundreds of millions of users worldwide~\cite{barowy2014excellint,kandel2011wrangler}. They empower users—often without formal programming backgrounds—to manipulate and analyze data through formulas composed on a tabular grid. However, writing correct and robust formulas remains a significant challenge. Small mistakes such as incorrect cell references, missing arguments, or improper function nesting can break computations or lead to incorrect results. These errors may surface as syntax problems, logical bugs, or runtime failures (e.g., \texttt{\#DIV/0!}, \texttt{\#REF!}), and diagnosing them is often non-trivial for non-programmers~\cite{hermans2016detecting}.
This highlights the need for tools that assist users in automatically detecting and fixing errors in their formulas.

% \begin{figure}
%     \centering
%     \includegraphics[width=1\linewidth]{figures/data_point_excel_benchmark.png}
%     \caption{Formula Repair Task and Dataset: Given a user-entered formula that produces a runtime error, the goal is to generate a corrected version that resolves the error and aligns with the user's intent. Intent is primarily inferred from the surrounding table, and optionally from a user utterance.}
%     \label{fig:example_data_point}
% \end{figure}

Automated Program Repair (APR) has been extensively studied for general-purpose programming languages (GPLs) like Java, Python, and C++~\cite{monperrus2018automatic,gousios2015work}. These systems typically rely on well-structured code with modular functions and comprehensive test suites, which support static analysis and test-driven repair strategies.
In contrast, spreadsheet formulas present unique challenges. They operate over structured data objects (e.g., cell ranges, tables) and often lack modular abstractions or formal specifications. Writing correct formulas requires not just syntactic fluency but also a clear understanding of how to reference and manipulate tabular data. Errors often arise from misinterpreting layouts, incorrect argument choices, or misunderstanding function semantics.

Prior work on APR in spreadsheets (henceforth referred to as {\em Excel formula repair}) has largely focused on syntactic repair, often without leveraging the spreadsheet context. For example, LaMirage~\cite{10.1145/3563327} uses grammar-based candidate generation and a neural ranker to fix syntax errors. A more recent system, RING~\cite{bavishi2022neurosymbolic} applies prompting techniques and error localization to formula text to suggest potential repairs to formula errors.
FLAME~\cite{joshi2024flame}, a domain-specific fine-tuned model has shown impressive performance on formula synthesis using compact transformers trained on Excel-specific corpora. 
However, none of these approaches consider tabular data. Also, most of these approaches remain focused on syntax repair, not semantic correctness.

Importantly, our analysis of real-world user queries on forums like Stack Overflow and Excel support channels reveals that a majority of formula issues are semantic rather than syntactic. These issues often manifest as runtime errors and require context-aware reasoning over both formula logic and spreadsheet data for effective resolution. 
Current Excel formula repair approaches do not address runtime errors. 
Moreover, existing datasets typically include only isolated formula pairs—incorrect and corrected—but omit the spreadsheet context necessary for modeling these semantic repair tasks.
There isn't an existing dataset that can be used to train and evaluate a repair model for runtime errors.

% \vspace{0.5em}
\noindent\textbf{Our Contributions.} To address this gap, we introduce \textsc{FoRepBench} (\textbf{Fo}rmula \textbf{Rep}air \textbf{Bench}mark), a new benchmark and dataset for context-aware Excel formula repair. Our main technical contributions are:
\vspace{-6pt}
\begin{enumerate}
    \item \textbf{Excel Formula Repair Dataset}: We present \forepbench{},  the first large-scale dataset of Excel formula repair examples for runtime errors. Each example includes spreadsheet context (cell values, headers), a broken formula, its corrected version, and a user utterance expressing intent.
    The dataset contains 618 examples and spans 5 runtime error types: \#DIV/0!, \#N/A, \#NAME?, \#REF!, and \#VALUE!.
    % The dataset spans a diverse range of errors, including syntactic bugs, logical faults, and runtime exceptions.
    
    \item \textbf{Synthetic Data Generation and Validation Pipeline}: We introduce a data generation pipeline that bootstraps from a small number of high-quality samples and produces realistic examples for training and evaluating models for Excel formula repair.
    % that bootstraps from correct formulas and applies targeted perturbations to simulate user errors. 
    We validate each repaired formula both via execution (to confirm correctness) and through a chain-of-thought LLM judge (to ensure semantic alignment with intent), resulting in high-quality examples.

    \item \textbf{Baseline Approach for Excel Formula Repair}: We propose a  method that leverages a large language model and incorporates both formula text and spreadsheet context for error correction. We also report the performance of the baseline approach on \forepbench{} and seed dataset using various state-of-art proprietary and open-source models.
    
    % our approach with four different LLM backbones on the Excel Formula Repair Task.
\end{enumerate}

We have made \forepbench{} available as a resource for further research on Excel formula repair and related tasks. \footnote{\url{https://github.com/microsoft/prose-benchmarks/tree/main/FoRepBench}}.

\section{Related Work}
\subsection{LLMs for Code Generation}
Large Language Models (LLMs) have emerged as a powerful paradigm for code generation. Early work such as GPT-3~\cite{brown2020language} showed that scaling autoregressive transformers in a few-shot setting can yield impressive results in synthesizing code from natural language prompts. This breakthrough spurred the development of models fine-tuned specifically on code, substantially improving both fluency and correctness. OpenAI's Codex~\cite{chen2021evaluating} adapts the GPT architecture with fine-tuning on a massive corpus of public code repositories and supports a wide range of programming tasks—from simple completions to complex algorithmic problems. Salesforce’s CodeGen~\cite{nijkamp2022codegen} and Meta AI’s InCoder~\cite{inCoder2022} introduced novel pretraining objectives, including span-masking and infilling, enabling the generation of code that integrates naturally within surrounding context.

Hybrid approaches have also been explored. DeepMind’s AlphaCode~\cite{li2022competition} combines LLMs with search-based techniques, achieving competitive results on programming competitions. Other models, like CodeT5~\cite{wang2021codet5}, leverage structural information by incorporating representations of abstract syntax trees (ASTs)~\cite{yin2018learning}, improving syntactic accuracy and semantic consistency. In our experiments, we evaluate our formula repair approach using four recent LLMs as the backbone: GPT-4o \cite{hurst2024gpt}, GPT-4.1 \footnote{https://openai.com/index/gpt-4-1/}, Phi-3 \cite{abdin2024phi}, and Mistral \cite{jiang2023mistral7b}, through prompt engineering tailored to the Excel formula repair task.

\subsection{Excel Formula Generation and Repair}
Research on code generation and repair in the context of Excel formulas remains relatively limited. One major line of work focuses on the NL-to-Formula (NL2F) task, which adapts the Text2SQL paradigm to Excel formula generation~\cite{zhao2024nl2formulageneratingspreadsheetformulas}. SpreadsheetCoder~\cite{chen2021spreadsheetcoderformulapredictionsemistructured} enhances formula prediction by incorporating spreadsheet context, improving the accuracy of generated formulas. FlashFill~\cite{10.1145/1926385.1926423} pioneered example-driven formula synthesis, enabling users to generate formulas via input-output examples. LaMirage~\cite{10.1145/3563327} targets the “last-mile” repair problem by fixing near-correct formulas using symbolic and neural techniques, but it does not leverage the surrounding spreadsheet data for deeper semantic reasoning.

Recent work such as FLAME~\cite{joshi2024flame} introduced a lightweight transformer model for formula completion and repair, trained specifically on Excel formulas. While effective, FLAME operates solely on formula syntax  and does not incorporate the spreadsheet context and natural language input, limiting its applicability in user-facing or intent-driven tasks. In contrast, our dataset includes natural language utterances alongside spreadsheet context, faulty formulas, and ground-truth repairs. This enables the study of broader problem settings—ranging from NL-to-formula generation to semantic formula repair—not just syntactic correction or last-mile fixes.

Using LLM-as-a-judge for synthetic data evaluation has emerged as an active research area across text generation~\cite{gudibande2023false,wang2023chatgpt,liu2023gpteval} and code generation tasks~\cite{zheng2023judging,chen2023codellm}.
Complementary to our work, Singh et al.~\cite{singh2024empiricalstudyvalidatingsynthetic} proposed an automated method for validating synthetic NL-to-formula datasets using LLMs. Their pipeline classifies and filters low-quality synthetic annotations to improve fine-tuning performance. We adopt a similar idea in our generation pipeline by incorporating LLM-based validation to ensure both execution correctness and semantic fidelity of generated examples.

\begin{figure*}
    \centering
    \includegraphics[width=0.65\textwidth]
    {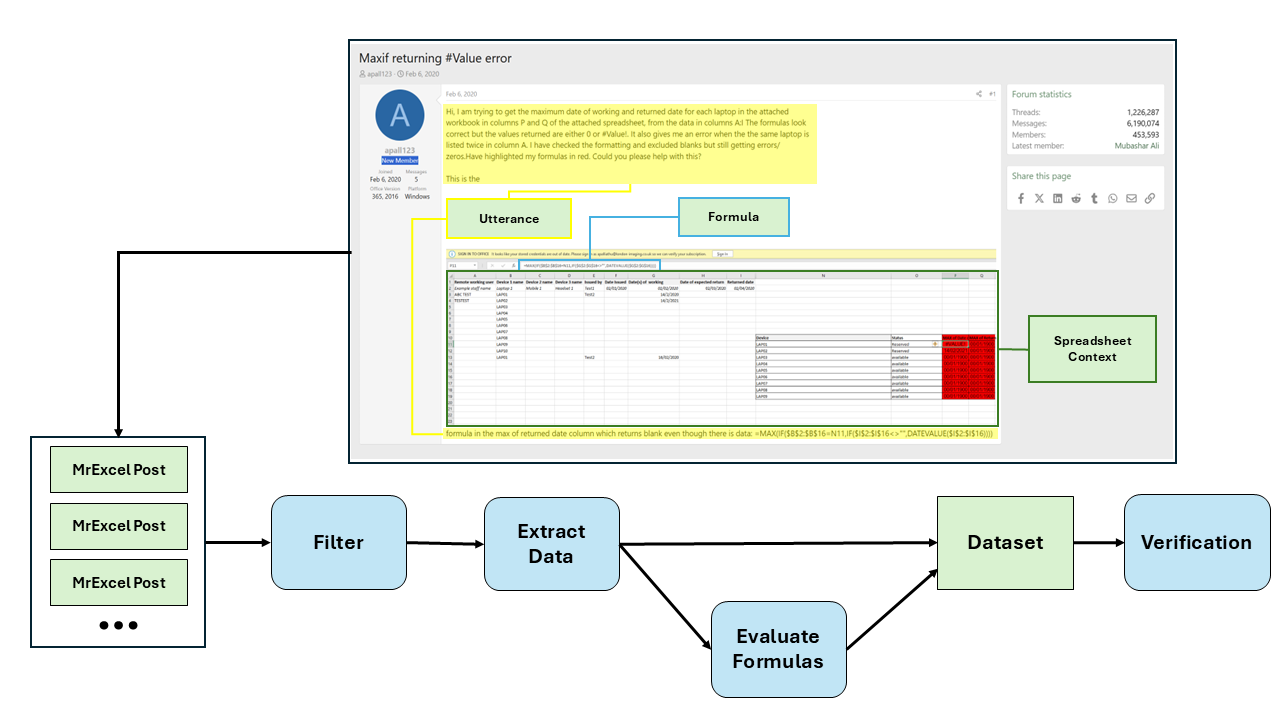}
    \caption{Overall workflow illustrating the sequential steps in the seed data curation process. MrExcel forum is scraped to get posts. Filters are used to identify posts containing table context, faulty formula and correct formula. The page is parsed to extract the required data. The faulty and correct formula are evaluated using Calc.ts. Samples where they execute as expected are added to the dataset. Finally the dataset is manually verified.}
    \label{fig:crawling_pipeline}
\end{figure*}

\section{Methodology}
This section describes our proposed methodology for synthetic benchmark generation, which  we refer to as \bootstrapgen{}.
Each data point in the benchmark, which represents a formula repair scenario, must include the following fields:
\begin{enumerate}
    \item \textbf{Tabular Data}: The spreadsheet context where the user encountered a runtime error.
    \item \textbf{Faulty Formula}: A formula that results in a runtime error. In this work, we focus on \#N/A, \#REF!, \#VALUE!, \#NAME?, and \#DIV/0! errors.
    \item \textbf{Correct Formula}: A formula that resolves the runtime error and is also consistent with the user intent expressed in the utterance.
    \item \textbf{Utterance}: A natural language query describing the user's problem and/or task that they are attempting to solve with their formula.
\end{enumerate}
(Refer to Figure \ref{fig:seed_data_example} for an example data point for Excel formula repair.)

Our data generation method relies on a small set of high quality examples to generate a larger dataset. In Section \ref{sec:seed_data_curation}, we describe how we curated a set of seed samples. Then in Section \ref{sec:BootstrapGeneration}, we describe our synthetic data generation approach which creates \forepbench{}.

\subsection{Seed Data Curation} 
\label{sec:seed_data_curation}
% \todo{Nick: Details of XL2BB crawling + Using LLM to summarize posts?}
To create a seed dataset for bootstrap generation, we developed a systematic approach to collect and process data from online forums where users discuss Excel-related problems and solutions. This section describes the methodology used to gather relevant seed data and prepare it for use in our synthetic data generation pipeline.

\subsubsection{Dataset Creation}
\label{sec:dataset_creation}
Figure \ref{fig:crawling_pipeline} shows an overview of our seed dataset creation approach.
We scraped posts from the MrExcel\footnote{https://www.mrexcel.com/} forum, a well-established platform where users frequently seek assistance with Excel formulas and share solutions. 
% An automated script was used to navigate through multiple pages of the forum. For each page, the script retrieved the HTML content, identified threads containing discussions about Excel formulas, and extracted the title, content, and all the replies for each thread.
Next, we filtered posts to ensure that they had all the required elements for our dataset, i.e. a faulty formula, table context, and the correct formula (we used a reply being marked as ``accepted answer'' on the forum as an indicator).
Once relevant posts were identified, we extracted the table context and formulas. Users often share tabular data and formulas in various formats, including plain text, code blocks, or specialized markup. We employed parsing techniques to accurately extract the exact text of the formulas used in cells, the data values in cells which may be referenced by the formulas, and any error messages or codes displayed by Excel, as reported by the users. Using the extracted data, we reconstructed the structure of the Excel workbooks by identifying different worksheets mentioned in the posts, mapping formulas and values to their corresponding cell addresses, and understanding the relationships between cells, such as which cells are referenced by a formula.
Since we want to create a dataset focused on runtime errors, we needed to identify posts where the faulty formula resulted in a runtime error. We simulated the evaluation of the extracted formulas using Calc.ts\footnote{\href{https://www.microsoft.com/en-us/garage/wall-of-fame/calc-ts-in-excel-for-the-web/?msockid=38b38871134a6f5806f59df512676e0c}{https://www.microsoft.com/en-us/garage/wall-of-fame/calc-ts-in-excel-for-the-web/?msockid=38b38871134a6f5806f59df512676e0c}}, an Excel formula evaluation engine capable of interpreting and calculating the results of formulas outside of the Excel application environment. This allowed us to detect the type of errors produced by the faulty formulas and test the corrected formulas to confirm that they resolve the errors. We retained posts for 5 runtime error types:  \#N/A, \#REF!, \#VALUE!, \#NAME?, and \#DIV/0!. For every post that passed, we added one sample to our dataset containing: 1) faulty formula, 2) correct formula, 3) table context, 4) user query, 5) runtime error type, and other metadata.

\subsubsection{Manual Verification and Correction}
Although we applied several automated filtering and validation steps, not all samples met our requirements for high-quality seed samples. This is because 1) a formula  that was accepted by a user on the forum as a solution and did not result in a runtime error when executed through Calc.ts could still be {\em semantically} incorrect, and 2) The table extracted by our scripts could contain data that is not part of the user's intended table context.  For example, in one of the samples, a column with the expected outputs was included in the context. A good benchmark sample should test a model's formula repair capability without leaking information. 

To address these limitations, we conducted two rounds of reviews to verify and correct the seed data. In the first round, each sample was assigned to one annotator. The annotator was asked to comment on the following: 
\begin{enumerate}
    \item Does the correct formula meet the requirements expressed in the utterance? 
    \item Does the faulty formula produce the required error type? 
    \item Is the table accurately extracted from the post?
    \item Is the table consistent with the user utterance?
\end{enumerate}

Only a third of the samples annotated in the first round met all the requirements. We therefore had a second round of annotations to verify the labels in the first round, and if necessary, edit the samples.
In this  round, the samples were reviewed again by a group of three or more annotators. If they agreed that the sample met the above requirements, it was added to the final seed dataset. If not, the sample was edited based on the comments from the first round. In some instances, the user utterance was edited to make it more explicit. In some cases, the formula crawled automatically was not correct for the user’s intent and was manually replaced with a truly correct formula. Some examples were deleted if they were too ambiguous or if the error was too trivial. 
This process ensured that for the samples in the seed dataset,  
\begin{enumerate}
    \item The table contains the necessary information for the repair, and not rows with the desired output that might leak information.
    \item The user utterance clearly indicates what the user wants to accomplish. 
\end{enumerate}

\subsection{Bootstrap Generation}
\label{sec:BootstrapGeneration}
\begin{figure*}
    \centering
    \includegraphics[width=0.65\textwidth]
    {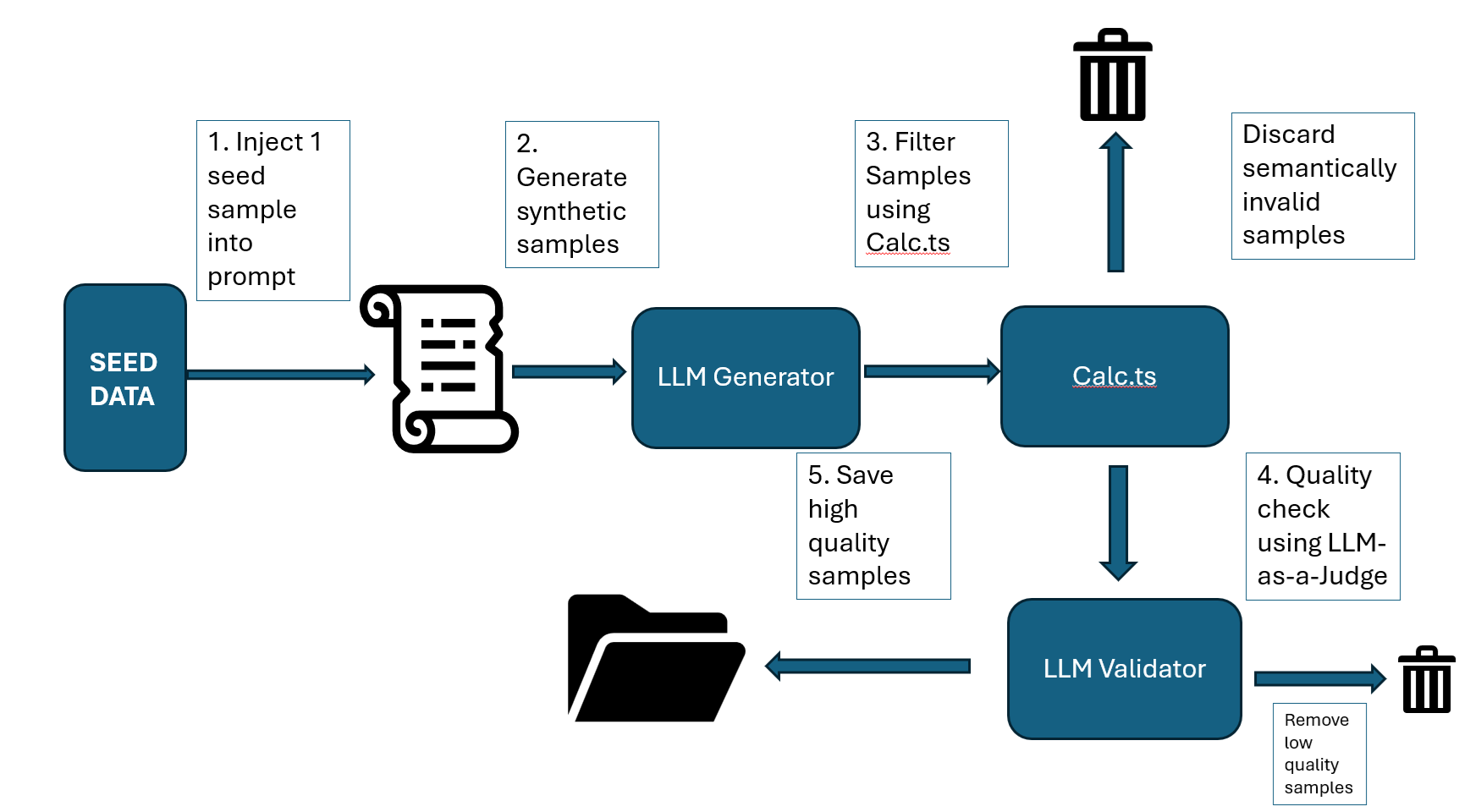}
    \caption{Pipeline overview of the synthetic data generation system.}
    \label{fig:synthetic_pipeline}
\end{figure*}

Automatic extraction of data from forums as discussed in Section \ref{sec:dataset_creation} led to incomplete and inaccurate samples, and manual validation and correction is not scalable to a large number of samples. In this section, we introduce our \bootstrapgen{} approach, using which we created \forepbench{}, a large scale benchmark dataset for Excel formula repair focused on runtime errors. Figure \ref{fig:synthetic_pipeline} shows an overview of the pipeline.
We start with a small number of high-quality samples, i.e. seed samples (See Section \ref{sec:seed_data_curation}), and generate a larger benchmark dataset synthetically.

The development of the synthetic data generation pipeline was guided by the following primary objectives:
\begin{enumerate}
    \item Ensure proper formatting of all synthetic data.
    \item Verify the correct execution of all synthetic formulas in Excel.
    \item Ensure that the data is semantically consistent
    \item Assess the quality of the data to ensure appropriate difficulty levels, function coverage, etc.
\end{enumerate}

The first step in the pipeline involves generating synthetic samples using few-shot prompting. 
Following the generation phase, it is essential to validate the synthesized samples to ensure both correctness and quality. 
This led to the creation of \forepbench{}, constructed through the following two-stage validation process. 
The complete pipeline consists of these three steps, as described below.
% The first step in the pipeline is generating synthetic samples through few-shot prompting.
% Following the generation of synthetic data, it was imperative to validate the samples to ensure both correctness and quality leading to creation of \forepbench{}, which was done in the following two steps. All three steps are described below.
% Correctness was assessed by verifying that the formulas executed correctly on the associated tables, while quality was evaluated to ensure the generated data met high standards.
% The remainder of this section elaborates on the prompting strategy employed to generate the synthetic data, followed by the methodologies used to evaluate the correctness and quality of the data.

\subsubsection{\textbf{Data Generation with Few-Shot Prompting}}
We apply \textit{one-shot} prompting to generate synthetic samples utilizing every sample from our seed data. This approach involved injecting each data point from our seed data into its own text prompt and subsequently generating new data based on that prompt. This method proved to be the most effective in our validation experiments which have been discussed in Section 3.1.

\subsubsection*{Fewshot-learning Setup Optimization}
To generate synthetic samples, we initially employed a zero-shot prompt to evaluate the LLM's capability to generate samples without grounding data. We generated 125 data points per error type, and the key observations were as follows:

\begin{enumerate}
    \item \textbf{Simple Data:} The generated Excel formulas and tables were relatively simple, e.g., dividing an arbitrary cell value by 0 to produce a \#DIV/0! error.
    \item \textbf{Lack of Diversity:} The resulting synthetic data exhibited minimal semantic and syntactic diversity. For a given error type, the model predominantly produced data points with minor differences to the table data or variables in the formula.
\end{enumerate}

Despite extensive prompt modifications to address the aforementioned issues, the problems persisted. Consequently, the next phase involved incorporating real-world grounding data into our prompts as few-shot examples for generating new data.

Next, we explored few-shot prompting as a more promising approach for creating a dataset with more diversity and complex examples. Our validation experiments showed that 1-shot prompting produced more samples that passed our validation tests (\S \ref{sec:calc_validator} and \S \ref{sec:llm_validator}) compared to zero-shot. Each data point from the seed dataset was used to produce multiple new samples.  
% This involved randomly selecting three seed data points corresponding to a specific error type as grounding data for the LLM. The quality of the generations was significantly better than the previous generations obtained from 0-shot prompting but the key issue was the number of samples generated. We generated fewer than 1/4 of the generations from the 0-shot prompting. To remedy this we used a third generation strategy: one shot prompting using every seed sample in our dataset.

\subsubsection{\textbf{Validating Generations executing Excel formulas}}
\label{sec:calc_validator}
To verify correctness, we utilized a tool called Calc.ts\footnote{\url{https://www.microsoft.com/garage/wall-of-fame/calc-ts-in-excel-for-the-web/}}, designed to check Excel formulas against the corresponding spreadsheet data. We ensured that each generated ``correct'' formula did not result in an error and that faulty formulas produced the appropriate runtime error. After confirming correctness, we evaluated the quality of the generated data using an LLM-as-a-judge framework. Samples that passed both correctness and quality checks were subsequently added to the final dataset. 
\subsubsection{\textbf{Validating Generations with LLM-as-a-Judge Approach}}
\label{sec:llm_validator}
To ensure reliability of the generated synthetic data, the LLM-judge approach we implemented leverages Chain-of-Though (CoT) reasoning for systematic assessment. We refer to this model as \llmvalidator{}. In this work, the LLM-judge systematically analyzes each repaired formula by first determining whether it resolves the original runtime error in the synthetic data. It then assesses whether the formula aligns with the user's intent, considering the spreadsheet context and any provided utterance. It is also prompted to assess and annotate the difficulty level of the repair, which we use for analysis of our proposed benchmark in Section \ref{RQ1}.

\section{Formula Repair}
As discussed in Section~\ref{sec:introduction},
there has been scarcity of research done on systems capable of Excel formula repair for formulas that result in semantic errors, rather than merely correcting syntax mistakes. 
% which is also tightly coupled with the data present within the sheets on which the programs operate.
A key challenge in Excel formula repair lies in incorporating the relevant spreadsheet context. Unlike general-purpose programming languages, Excel formulas are tightly coupled to tabular data layouts, and the correct repair often depends on values, ranges, headers, or even user-entered text elsewhere in the spreadsheet.

To address this, we propose a baseline solution that not only leverages the buggy formula and any available auxiliary information (such as natural language descriptions), but also uses the context present within the spreadsheet.
% a context retrieval component, that feeds information about the tabular data present within the spreadsheet.
Our system thereby enables context-aware formula repair that handles both syntactic and semantic errors. We further utilize our baseline repair pipeline to evaluate \forepbench{}, thereby demonstrating its practical relevance and showcasing how such controlled benchmarks can effectively approximate real-life formula repair scenarios encountered in production spreadsheets.

% Automated program repair has been an active area of research, where the goal is to automatically fix broken code that results in errors when compiled/executed. While significant progress has been made in repairing programs written in high-resource languages such as Python and Java — addressing both syntactic and semantic errors — low-resource languages like Spreadsheet Formulas and PowerQuery remain under-explored. In particular, there has been scarcity of research done on systems capable of repairing formulas that result in semantic errors, rather than merely correcting syntax mistakes.

% A key challenge in formula repair, especially for semantic errors- lies in correctly identifying and incorporating the relevant spreadsheet context. Unlike general-purpose programming languages, spreadsheet formulas are tightly coupled to tabular data layouts, and the correct repair often depends on values, ranges, headers, or even user-entered text elsewhere in the spreadsheet. However, due to the limited input token budget of large language models, it is impractical to simply provide the entire spreadsheet as input during repair.

\subsection{Baseline Repair Technique}
\label{section:rep solu}
The baseline method follows a structured pipeline that makes a single call to an LLM, designed to efficiently process faulty Excel formulas and generate repaired versions along with explanations. Figure~\ref{fig:synthetic_pipeline} illustrates the single-call LLM solution used for evaluating the benchmark. The system takes as input a faulty formula, the corresponding runtime error, and an optional user utterance. Since spreadsheet tables can be large, passing the entire spreadsheet as context is impractical due to LLM token limitations. To address this, we identify the nearest table associated with the faulty formula and extract its header along with a few sample rows to provide as context. This enables the LLM to receive the necessary contextual information to generate accurate repairs.

Once the relevant spreadsheet context is retrieved, a structured prompt is constructed. This prompt consists of four key elements: the extracted spreadsheet data context, the faulty formula, the runtime error, and an optional user utterance (if provided). In this work, a standardized prompt template is used to maintain consistency, and it included instructions that guide the LLM in repairing Excel formulas while ensuring a meaningful explanation is generated. The constructed prompt is then passed to the LLM, which processes the input and attempts to generate a corrected formula along with a natural language explanation of the fix. The repaired formula is subsequently evaluated by comparing it with the ground truth correct formula from the benchmark dataset. This comparison helps determine whether the generated repair successfully resolves the runtime error while maintaining the intended logic of the original formula. This structured pipeline used in this work provides a consistent and repeatable methodology for evaluating formula repair performance.
% we're not including stuff about metaprompt design
% More details about the meta prompt design are available in the Appendix [\textcolor{red}{Emmanuel -- Include prompt template in appendix}].

\begin{figure}
    \centering
    \includegraphics[width=1\linewidth]{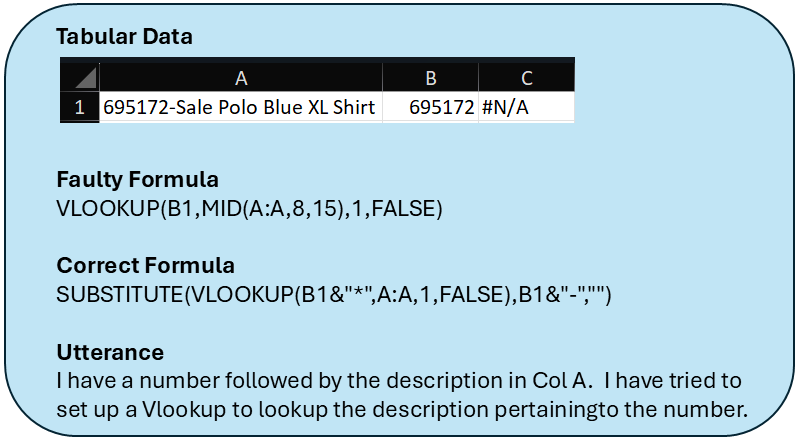}
    \caption{An example from the seed dataset where the faulty formula applies substring extraction, while the correct formula uses wildcard matching and string substitution—requiring multiple semantic edits to fix.}
    % An example from the seed dataset, where the user wants to extract a part of a string from a cell. The faulty formula has a wrong implementation of the substring extraction which is very far away from the correct formula. }
    \label{fig:seed_data_example}
\end{figure}

\begin{figure}
    \centering
    \includegraphics[width=\linewidth]{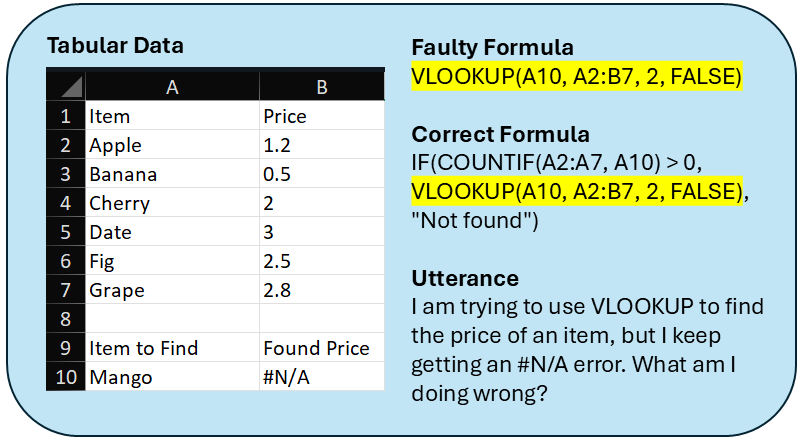}
    \caption{An example from \forepbench{}. The faulty formula does not handle the exception when the value A10 is not present in the table. The correct formula fixes this by checking if the value exists, and returning "Not Found" if it doesn't.}
    \label{fig:synthetic_data_example}
    \vspace{-9pt}
\end{figure}

\section{Experimental Setup}
We applied \bootstrapgen (Section \ref{sec:BootstrapGeneration}) to generate our dataset using \texttt{GPT-4o} as the LLM. A temperature of 0.64 is used to promote diversity among generated examples. In total, we generated 1095 samples, out of which 618 passed \llmvalidator{}. Table \ref{tab:llm_validator} shows the error-wise breakdown.
We conducted analysis to assess the characteristics and quality of our dataset. We address the following research questions:
\begin{rqlist}
    \item Does the data distribution in the \forepbench{} match real world data? How does the generated data look in terms of distribution across errors and functions?
    % \todo[inline]{the experiment measures the correctness of the LLM Validator used in our dataset generation pipeline, maybe that should be the RQ as well}
    \item What is the quality of the \forepbench{} based on human judgments?
    
    \item What is the performance of the proposed baseline repair approach  across a range of state-of-the-art proprietary ( i.e, \texttt{Gpt-4.1}, \texttt{Gpt-4o}) and open-source LLMs (i.e, \texttt{Phi-3}, \texttt{Mistral}) on \forepbench{}?
    
    \item What is the cost of generating \forepbench{}? How many LLM calls are needed?
\end{rqlist}

To answer RQ1, we present the difficulty and function distributions for both the seed dataset and \forepbench{}. The difficulty of each sample was determined by \llmvalidator{}. It was prompted to assign one of 3 difficulty ratings to the sample based on how complex the Excel repair task was - easy, medium, and hard. 

To answer RQ2, we recruited 2 annotators to assess the quality of the generated data {\em before} it was passed through \llmvalidator{}. Due to the complexity of the task and limited resources, two teammates with extensive familiarity with the task and deeper understanding of the nuances served as annotators in this study. They annotated a subset of 24 samples from the synthetic dataset. They were asked to perform the same task as \llmvalidator{} in Section~\ref{sec:BootstrapGeneration}, i.e. check the correctness and consistency of the table context, faulty formula, correct formula, and utterance.

\subsection{Metrics}
\label{sub_Sec:metrics}
To evaluate the performance of the baseline repair technique across datasets, we employ the following execution-based metrics:
% \vspace{-9pt}
\paragraph{\textbf{Syntax Validity:}} We first check whether the repaired Excel formula can be successfully compiled. If the formula parses without any compilation errors, it is considered syntactically valid; otherwise, it is marked as invalid.
% \vsp/ace{-9pt}
\paragraph{\textbf{Can Execute:}} This metric verifies whether the repaired formula can be successfully executed on the spreadsheet without triggering any runtime errors (e.g., \texttt{\#VALUE!}, \texttt{\#REF!}, \texttt{\#DIV/0!}, etc.). A successful execution without runtime errors is considered a success; otherwise, it is considered a failure.
% \vspace{-9pt}
\paragraph{\textbf{Execution Match:}} After execution, we compare the output produced by the repaired formula against the output of the ground-truth (correct) formula. If the outputs match exactly the repair is considered correct under this metric.

 \begin{figure}
    \centering
    \includegraphics[width=\linewidth]{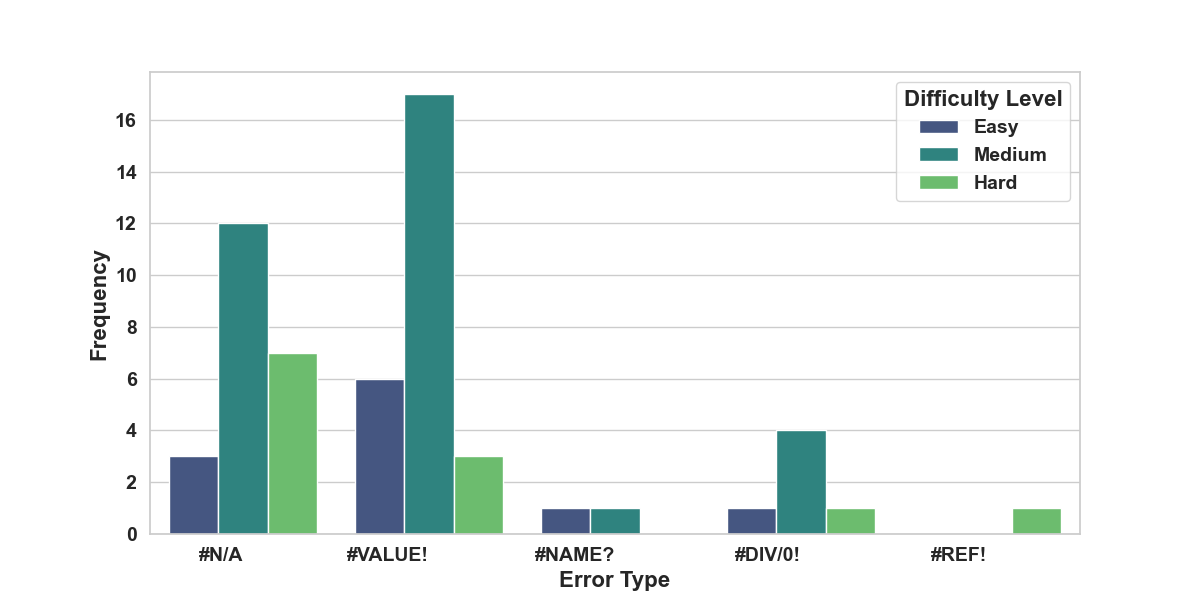}
    \caption{Distribution of sample difficulty levels by Excel error type in the seed dataset. Difficulty levels were assigned by \llmvalidator{}.}
    \label{fig:difficulty_dist_seed}
\end{figure}
% \vspace{-19pt}
\begin{figure}
    \centering
    \includegraphics[width=\linewidth]{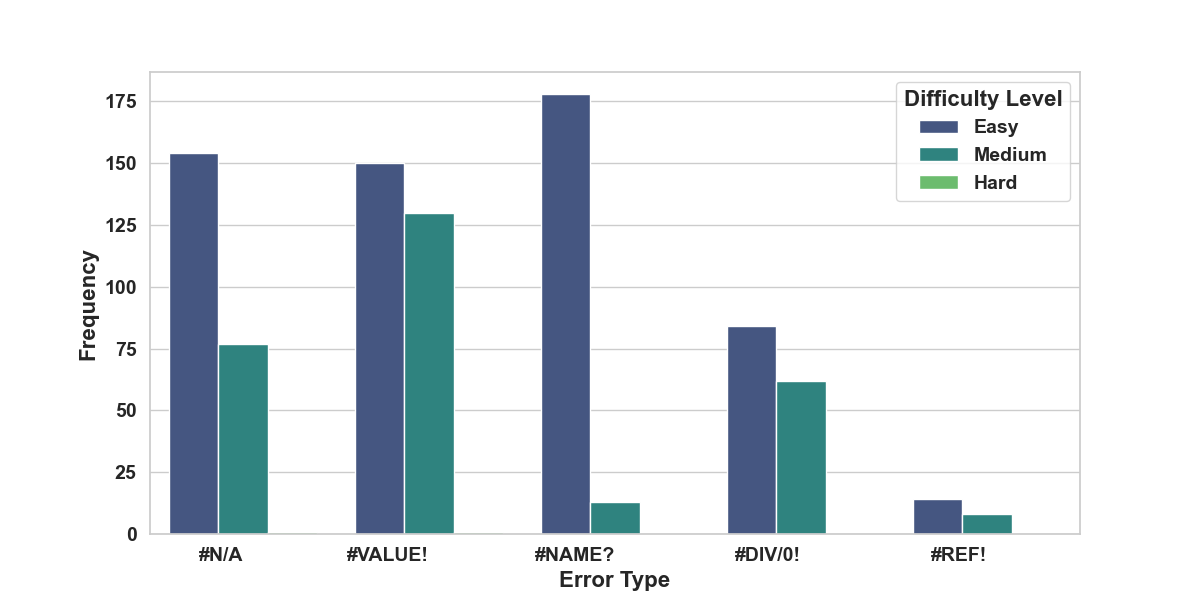}
    \caption{Distribution of sample difficulty levels by Excel error type in \forepbench{}. Difficulty levels were assigned by \llmvalidator{}.}
    \label{fig:difficulty_dist_synthetic}
\end{figure}

\begin{table}[h]
\begin{tabular}{@{}lcc@{}}
\toprule
Error Type & Samples Generated & Samples passed by \llmvalidator{}   \\    \midrule
\#VALUE!            & 241                    & 64                     \\
\#N/A               & 329                    & 140                    \\
\#REF!              & 16                     & 12                     \\
\#NAME?             & 222                    & 158                    \\
\#DIV/0!            & 287                    & 244                
\\ \bottomrule
\end{tabular}
\caption{The table shows the number of generated samples that remained after \llmvalidator{} filtered out low quality samples. Overall, 56\% of all generated samples pass \llmvalidator{}.}
\label{tab:llm_validator}
\vspace{-29pt}
\end{table}

% \vspace{-19pt}
\section{Results}
\subsection{RQ1: Data Distribution and Comparison with Seed Dataset}
\label{RQ1}
Figures~\ref{fig:difficulty_dist_seed} and~\ref{fig:difficulty_dist_synthetic} display the distribution of difficulty levels across error types in the seed dataset and \forepbench{}, respectively. Compared to the seed dataset, the samples in \forepbench{} are skewed toward simpler formula repairs. This suggests that \bootstrapgen{} tends to produce less complex scenarios, as it's hard to synthesize examples that both execute successfully and pass the LLM-based quality filter \llmvalidator{}. Figures \ref{fig:seed_data_example} and \ref{fig:synthetic_data_example} illustrate this disparity in complexity. Both examples have faulty formulas with the VLOOKUP function. In the example from the seed dataset in Figure \ref{fig:seed_data_example}, the user wants to extract a part of a string from a cell, a semantically complex scenario. In a simpler (more syntactic) scenario from \forepbench{} shown in Figure \ref{fig:synthetic_data_example}, the user only needs to handle an exception when the value being looked up isn't present in that cell range.

Tables ~\ref{tab:function_distribution_seed} and ~\ref{tab:function_distribution} present the distribution of Excel functions in the seed dataset and \forepbench{} respectively, broken down by error type. We include the top 10 functions that appear in the most examples in this dataset.
% The most common functions—such as \texttt{SUM}, \texttt{INDEX}, \texttt{MATCH}, and \texttt{VLOOKUP}, appear frequently in association with errors like \#DIV/0! and \#N/A. These patterns suggest a strong correlation between certain functions and the types of formula errors observed in synthetic samples. 
Comparing the distribution with Table \ref{tab:function_distribution_seed}, \forepbench{}  frequently has the functions AVERAGE and CONCATENATE which the seed dataset did not. This indicates that \bootstrapgen{} creates samples diverse from the fewshot examples. 
% Also, it contains "AVRG", which is not an actual Excel function and is used in examples of \#NAME? errors. 
Some trends remain consistent between both datasets, for example VLOOKUP, INDEX, and MATCH being the most common functions for \#N/A error.
% Although it includes a more diverse set of functions, consistent trends emerge. For example, \texttt{VLOOKUP}, \texttt{MATCH}, and \texttt{INDEX} remain among the most frequent contributors to \#N/A and \#VALUE! errors. The seed dataset exhibits a broader functional variety and more complex combinations, which aligns with its higher repair difficulty.

% Table \ref{tab:formula_repair_results} shows the performance of our proposed solution on both the seed data and the synthetic data. The model performance is much better on the synthetic data, confirming that the generated samples are indeed easier formula repairs.

% Tables \ref{tab:function_distribution_seed} and \ref{tab:function_distribution} show the most common functions in each dataset, and how they are distributed across the 5 different error types. Certain trends remain common across both datasets. VLOOKUP, INDEX, and MATCH functions are among the most frequent in both datasets, and cause \#N/A error the most. 
\begin{table*}[ht]
\centering
\rowcolors{2}{white}{gray!10}
\begin{tabular}{lcccccccccccc}
\hline
Error Type & AND & FIND & IF & INDEX & LEFT & MATCH & MID & MIN & SUM & VLOOKUP \\ \hline
\#DIV/0! & \cellcolor[HTML]{FFFFFF} 0 & \cellcolor[HTML]{FFFFFF} 0 & \cellcolor[HTML]{95E9B5} 1 & \cellcolor[HTML]{FFFFFF} 0 & \cellcolor[HTML]{FFFFFF} 0 & \cellcolor[HTML]{FFFFFF} 0 & \cellcolor[HTML]{FFFFFF} 0 & \cellcolor[HTML]{2BD46C} 2 & \cellcolor[HTML]{95E9B5} 1 & \cellcolor[HTML]{FFFFFF} 0 \\
\#N/A & \cellcolor[HTML]{E9FAF0} 1 & \cellcolor[HTML]{FFFFFF} 0 & \cellcolor[HTML]{D4F6E1} 2 & \cellcolor[HTML]{55DC89} 8 & \cellcolor[HTML]{D4F6E1} 2 & \cellcolor[HTML]{2BD46C} 10 & \cellcolor[HTML]{E9FAF0} 1 & \cellcolor[HTML]{FFFFFF} 0 & \cellcolor[HTML]{E9FAF0} 1 & \cellcolor[HTML]{6AE098} 7 \\
\#NAME? & \cellcolor[HTML]{FFFFFF} 0 & \cellcolor[HTML]{FFFFFF} 0 & \cellcolor[HTML]{2BD46C} 1 & \cellcolor[HTML]{FFFFFF} 0 & \cellcolor[HTML]{2BD46C} 1 & \cellcolor[HTML]{FFFFFF} 0 & \cellcolor[HTML]{FFFFFF} 0 & \cellcolor[HTML]{FFFFFF} 0 & \cellcolor[HTML]{FFFFFF} 0 & \cellcolor[HTML]{FFFFFF} 0 \\
\#REF! & \cellcolor[HTML]{FFFFFF} 0 & \cellcolor[HTML]{FFFFFF} 0 & \cellcolor[HTML]{FFFFFF} 0 & \cellcolor[HTML]{FFFFFF} 0 & \cellcolor[HTML]{FFFFFF} 0 & \cellcolor[HTML]{FFFFFF} 0 & \cellcolor[HTML]{FFFFFF} 0 & \cellcolor[HTML]{FFFFFF} 0 & \cellcolor[HTML]{FFFFFF} 0 & \cellcolor[HTML]{FFFFFF} 0 \\
\#VALUE! & \cellcolor[HTML]{DBF7E6} 2 & \cellcolor[HTML]{B8F0CE} 4 & \cellcolor[HTML]{2BD46C} 12 & \cellcolor[HTML]{DBF7E6} 2 & \cellcolor[HTML]{DBF7E6} 2 & \cellcolor[HTML]{CAF4DA} 3 & \cellcolor[HTML]{CAF4DA} 3 & \cellcolor[HTML]{EDFBF2} 1 & \cellcolor[HTML]{DBF7E6} 2 & \cellcolor[HTML]{FFFFFF} 0 \\
\hline
\end{tabular}
\caption{This table reports the frequency of the top 10 most frequent functions in the seed dataset (N=59), split by error type. Darker color signifies higher frequency within each row.}
\label{tab:function_distribution_seed}
\vspace{-11pt}
\end{table*}

\begin{table*}[ht]
\centering
\rowcolors{2}{white}{gray!10}
\begin{tabular}{lcccccccccc}
\hline
Error Type & AVERAGE & AVRG & CONCATENATE & COUNT & IF & INDEX & MATCH & SUM & VALUE & VLOOKUP \\ \hline
\#DIV/0! & \cellcolor[HTML]{B8F0CE} 3 & \cellcolor[HTML]{FFFFFF} 0 & \cellcolor[HTML]{FFFFFF} 0 & \cellcolor[HTML]{A0EBBD} 4 & \cellcolor[HTML]{E7FAEE} 1 & \cellcolor[HTML]{FFFFFF} 0 & \cellcolor[HTML]{FFFFFF} 0 & \cellcolor[HTML]{2BD46C} 9 & \cellcolor[HTML]{FFFFFF} 0 & \cellcolor[HTML]{FFFFFF} 0 \\
\#N/A & \cellcolor[HTML]{FBFEFC} 1 & \cellcolor[HTML]{FFFFFF} 0 & \cellcolor[HTML]{FFFFFF} 0 & \cellcolor[HTML]{FFFFFF} 0 & \cellcolor[HTML]{FBFEFC} 1 & \cellcolor[HTML]{AEEEC7} 25 & \cellcolor[HTML]{A5ECC0} 28 & \cellcolor[HTML]{FBFEFC} 1 & \cellcolor[HTML]{FFFFFF} 0 & \cellcolor[HTML]{2BD46C} 66 \\
\#NAME? & \cellcolor[HTML]{95E9B5} 2 & \cellcolor[HTML]{2BD46C} 4 & \cellcolor[HTML]{FFFFFF} 0 & \cellcolor[HTML]{FFFFFF} 0 & \cellcolor[HTML]{95E9B5} 2 & \cellcolor[HTML]{FFFFFF} 0 & \cellcolor[HTML]{FFFFFF} 0 & \cellcolor[HTML]{CAF4DA} 1 & \cellcolor[HTML]{FFFFFF} 0 & \cellcolor[HTML]{FFFFFF} 0 \\
\#REF! & \cellcolor[HTML]{FFFFFF} 0 & \cellcolor[HTML]{FFFFFF} 0 & \cellcolor[HTML]{FFFFFF} 0 & \cellcolor[HTML]{FFFFFF} 0 & \cellcolor[HTML]{FFFFFF} 0 & \cellcolor[HTML]{FFFFFF} 0 & \cellcolor[HTML]{FFFFFF} 0 & \cellcolor[HTML]{95E9B5} 1 & \cellcolor[HTML]{FFFFFF} 0 & \cellcolor[HTML]{2BD46C} 2 \\
\#VALUE! & \cellcolor[HTML]{87E6AC} 9 & \cellcolor[HTML]{FFFFFF} 0 & \cellcolor[HTML]{2BD46C} 16 & \cellcolor[HTML]{FFFFFF} 0 & \cellcolor[HTML]{7AE4A3} 10 & \cellcolor[HTML]{E4F9EC} 2 & \cellcolor[HTML]{D7F6E3} 3 & \cellcolor[HTML]{6DE199} 11 & \cellcolor[HTML]{CAF4DA} 4 & \cellcolor[HTML]{FFFFFF} 0 \\
\hline
\end{tabular}
\caption{The table reports the frequency of the top 10 most frequent functions present in \forepbench{} (N=618), split by error type. Darker color signifies higher frequency within each row. Comparing the distribution with Table \ref{tab:function_distribution_seed}, \forepbench{}  frequently has the functions AVERAGE and CONCATENATE which the seed dataset did not. This indicates that \bootstrapgen{} creates samples diverse from the fewshot examples. Some trends remain consistent between both datasets, for example VLOOKUP, INDEX, and MATCH being the most common functions for \#N/A error.}
\label{tab:function_distribution}
\vspace{-13pt}
\end{table*}

\begin{figure}
    \centering
    \includegraphics[width=1\linewidth]{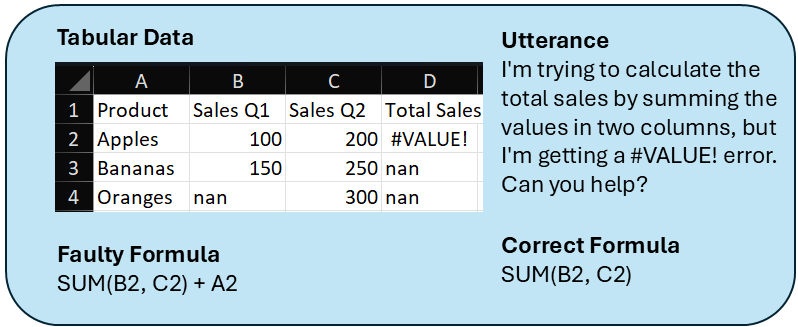}
    \caption{Example from the synthetic dataset {\em before} being passed through \llmvalidator{}.  where annotators disagree on validity. The formula includes cell \texttt{A2} in a summation, but \texttt{A2} contains a string. One annotator considers it valid, interpreting it as a possible typo, while the other marks it as invalid.}
    \label{fig:annotator_disagreement_example}
\end{figure}

\begin{figure}
    \centering
    \includegraphics[width=1\linewidth]{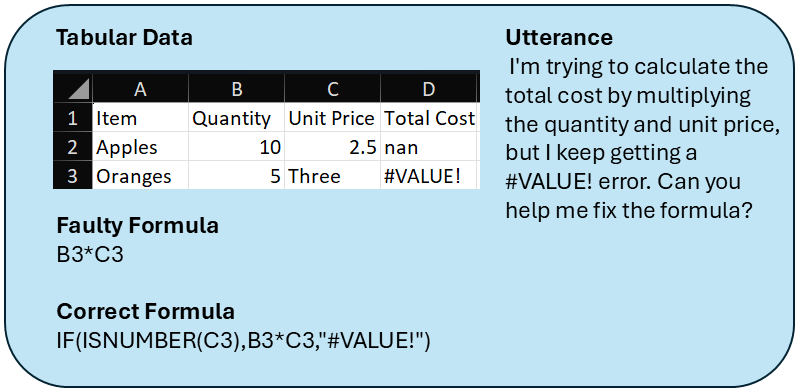}
    \caption{Example from \forepbench{} where human annotators disagree with \llmvalidator{} regarding validity. In this case, the "Unit Price" column contains the string "Three" instead of a numeric value. While annotators label the example as invalid, \llmvalidator{} incorrectly accepts it, highlighting its limitations in filtering out unrealistic inputs.}
    \label{fig:llm_disagreement_example}
\end{figure}

\subsection{RQ2: Synthetic Dataset Quality Based on Human Evaluation}
Table~\ref{tab:human_evaluation_results} reports Cohen’s Kappa scores measuring pairwise agreement between the two human annotators and between each annotator and \llmvalidator{}. The results indicate moderate agreement among human annotators (Kappa: 0.60), suggesting some subjectivity in evaluating the quality of generated samples. Annotators were asked to assess whether the corrected formula appropriately fixes the faulty one and satisfies the intended user operation, and whether the example reflects a realistic spreadsheet scenario. Disagreements often stemmed from differing interpretations of what constitutes a plausible Excel table. For instance, in the example in Figure \ref{fig:annotator_disagreement_example}, where a formula attempted to multiply a string by a number, one annotator interpreted it as a plausible user typo, while the other considered it unrealistic. This suggests the need for more concrete annotation guidelines with detailed examples for the annotators.

Agreement between \llmvalidator{} and each annotator was also moderate (0.42 for both annotators) but notably lower than inter-annotator agreement. In most cases, \llmvalidator{} accepted examples that were consistent in formula execution but unrealistic in context. For example, in the sample in Figure \ref{fig:llm_disagreement_example}, the table has textual values like “three” in a numeric column. While \llmvalidator{} deemed such examples valid based on logical consistency, human annotators rejected them due to implausible table semantics. These results imply that while \llmvalidator{} is effective at rejecting invalid or illogical formula pairs, it lacks sensitivity to the contextual plausibility of spreadsheet content. As a result, some unrealistic samples may persist in the dataset despite passing automated filtering.

\begin{table}[ht]
\centering
\begin{tabularx}{\linewidth}{lXXX}
\hline
          & Ann-1 vs. LLM & Ann-2 vs. LLM & Ann-1 vs. Ann-2 \\
\hline
Kappa & 0.42 & 0.42 & 0.60 \\
\hline
\end{tabularx}
\caption{The table reports Cohen’s Kappa agreement scores between human annotators and \llmvalidator{}.}
\label{tab:human_evaluation_results}
\vspace{-19pt}
\end{table}

% - Kappa table
% why the agreement was less . more learnings

\begin{table*}[]
\centering

\begin{tabular}{@{}llccc@{}}
\toprule
\textbf{LLM} & \textbf{Dataset Origin} & \textbf{Syntax Valid} & \textbf{Can Execute} & \textbf{Execution Match} \\
\midrule
\multirow{2}{*}{\textbf{GPT-4.1}} & \forepbench{} & \textbf{1.00} & \textbf{0.96}  & \textbf{0.80}  \\
                                  & Seed Dataset      & \textbf{0.98}      & \textbf{0.65}      & \textbf{0.41}      \\ \midrule
\multirow{2}{*}{\textbf{GPT-4o}}  & \forepbench{} & 1.00& 0.93& 0.73  \\
                                  & Seed Dataset      & 0.96 & 0.63 & 0.35 \\ \midrule
\multirow{2}{*}{\textbf{Phi-3}}   & \forepbench{} & 0.81  & 0.77  & 0.58  \\
                                  & Seed Dataset      &       0.73&      0.41 &       0.24\\ \midrule
\multirow{2}{*}{\textbf{Mistral}} & \forepbench{} & 0.78  & 0.76  & 0.51  \\
                                  & Seed Dataset      &       0.67&       0.37&  0.19\\    
                                  \bottomrule
\end{tabular}%

\caption{Performance of Baseline Repair Technique on \forepbench{} and seed dataset across various LLMs.}
\label{tab:baseline-repair-sol perf}
\vspace{-19pt}
\end{table*}

% \vspace{-19pt}
\subsection{RQ3: Performance of Repair Task on Synthetic and Seed Data}
% We evaluate the performance of the baseline Excel formula repair technique described in Section~\ref{section:rep solu} on both \forepbench{} and the seed dataset. 
% we chose a variety of 4 representative models considering aspects like state-of-the-art LLM, SLM, closed-source, open-source, transformer-based, and mixture-of-experts-based architectures to ensure experimental diversity
We evaluate the performance of the baseline excel formula repair technique described in Section~\ref{section:rep solu} on both \forepbench{} and the seed dataset. To ensure diversity in our evaluation, we selected four representative LLMs spanning a range of model families, including state-of-the-art proprietary large language models (i.e, \texttt{Gpt-4.1}, \texttt{Gpt-4o}) and open-weight models (i.e, \texttt{Phi-3}, \texttt{Mistral}), as well as different architectural paradigms (transformer-based and mixture-of-experts). 
% This selection allows us to assess the robustness of the repair technique across varied model capabilities and deployment scenarios.
Table~\ref{tab:baseline-repair-sol perf} reports the results across the three execution-based metrics described in Section~\ref{sub_Sec:metrics}, evaluated over both datasets using different LLMs that power the baseline repair technique.

Overall, we observe that both the \emph{Execution Match} and \emph{Can Execute} scores are significantly lower across LLMs on the seed dataset compared to \forepbench{}, indicating that the \forepbench{} is relatively easier for the repair technique to handle which matches our learning from the RQ~\ref{RQ1} on difficulty level distribution. Upon further analysis, we identify two primary reasons for this discrepancy:
(1) the faulty formulas in the seed dataset are generally more complex, often involving deeper levels of nesting; and
(2) the number of edits required to transform the faulty formula into the correct version is substantially higher.

For instance, consider a faulty formula from the manually annotated seed dataset shown in Figure~\ref{fig:seed_data_example}. This example contains nesting of depth two and uses both \texttt{VLOOKUP} and \texttt{MID} functions. To repair this formula, multiple non-trivial edits are required, including altering the internal logic from $MID(A:A, 8, 15)$ to simply $A:A$ within the \texttt{VLOOKUP} function, followed by additional transformations where the looked-up value is further modified to $B1 \& "-"$. Such repairs demand reasoning about user intent and understanding of higher-level semantics, as the modifications involve significant logic changes rather than isolated token-level corrections.

In contrast, examples from the synthetic dataset often require fewer edits to achieve the correct formula, as illustrated in Figure~\ref{fig:seed_data_example}. In these cases, the internal logic embedded within the \texttt{VLOOKUP} function typically remains intact, with only minor corrections or additional function wrap-up are needed. Consequently, these repairs are more straightforward for the model to handle, as they involve localized changes rather than substantial semantic rewrites. We also see that bigger GPT-4 series models are better at solving the repair tasks in comparison to Phi-3 and Mistral suggesting more layers and training data can help improve Excel formula repair task.

These findings suggest that while the baseline repair technique is effective on \forepbench{}, the \bootstrapgen{} pipeline may not fully capture the range of complexity observed in real-world Excel formulas. Specifically, the synthetic instances from \forepbench{} tend to exhibit shallower nesting and require fewer semantic transformations compared to the manually curated seed dataset. Building on this insight, we propose that incorporating an additional \textit{reviewer agent} or a \textit{human-in-the-loop component} into the \bootstrapgen{} pipeline could help bridge this gap. By comparing generated synthetic samples against real-world instances and providing targeted feedback, the pipeline could iteratively generate more complex and diverse faulty formula instances that better reflect real-world repair challenges.

\subsection{RQ4: Cost of Dataset Generation}
Table~\ref{tab:llm_calls} summarizes the cost of generating the \forepbench{} dataset in terms of LLM API usage. For data generation, we issue a single LLM call per prompt to generate $N$ candidate samples (where $N=25$). From this pool, 1,095 samples contained executable formulas with the correct error type, as automatically verified by Calc.ts. For each of these accepted samples, we issue one additional LLM call to assess semantic validity, resulting in a total of 1,154 LLM calls (59 for generation and 1,095 for validation). This corresponds to an average of approximately 2.09 LLM calls per accepted sample.

On average, each call consumes approximately 2,000 tokens, yielding a per-sample generation cost of approximately \$0.02\footnote{\url{https://llmpricecheck.com/openai/gpt-4o/}}. These results demonstrate that the \bootstrapgen{} pipeline is both scalable and cost-effective for generating large-scale formula repair datasets.

% Table~\ref{tab:llm_calls} summarizes the cost of generating the \forepbench{} dataset in terms of LLM API usage. To create the dataset, we use a single LLM call per prompt to generate $N$ candidate samples (here  $N=10$). From this set, 1,095 samples contain executable formulas with the correct error type, as verified by Calc.ts. For each of these samples, we make an additional LLM call to assess semantic validity. This results in a total of 1,154 LLM calls (59 for generation and 1,095 for validation), averaging approximately 2.09 calls per accepted sample.
% In an average each call has a token usage of 2k, which brings the cost of generating each sample to \$0.02, which shows that the \bootstrapgen{} is scalable ans cost effective.
% make one LLM call for each sample to generate N samples. As the temperature is set to 0.64, this generates a relatively diverse set of new samples. Out of all the generated samples, 1095 contain executable formulas with the right error type, as verified by Calc.ts. For each of these samples, we make another LLM call to assess validity. This leads to:

\begin{table}[ht]
\centering
\begin{tabular}{lll}
\hline
Generation Calls & Validation Calls & Avg. Calls/Valid Sample \\
\hline
59 & 1,095 & 2.09 \\
\hline
\end{tabular}
\caption{LLM call breakdown for generating the \forepbench{} dataset. Out of 1,095 executable samples, 618 are accepted by \llmvalidator{}.}
\label{tab:llm_calls}
\vspace{-11pt}
\end{table}

\vspace{-17pt}
\section{Discussion and Conclusion}
In this work, we introduced a modular, low-supervision pipeline for generating and validating synthetic Excel formula repair data, resulting in the \forepbench{} benchmark. Our \bootstrapgen{} method combines structured spreadsheet context with prompt-based LLM sampling to produce realistic faulty formulas, which are then filtered through a multi-stage validation process. This process leverages automated execution checks and an LLM-based reviewer agent that evaluates semantic plausibility. We also proposed a prompt-based repair baseline system to evaluate model performance on both synthetic and real (seed) data. Our investigation across four research questions uncovered key insights about the characteristics, challenges, and utility of the generated data.

From RQ1, we found the synthetic dataset covers a broad range of formula categories and function types, achieving greater diversity than the manually curated seed dataset. RQ2 reinforced this, showing that although synthetic data spans multiple function categories, its error types and logical transformations are more localized and less complex than those in seed data. RQ3 examined LLM-based repair performance on both datasets, revealing a stark contrast: models like GPT-4o and GPT-4.1 perform well on synthetic samples (execution match rates up to 0.80), but accuracy drops significantly on seed data, which often requires deeper nesting and multi-step reasoning. This gap highlights a limitation of synthetic generation-despite lexical and structural diversity, synthetic examples lack the semantic complexity of real-world formulas that demand substantial rewrites or ambiguous intent interpretation. Conversely, synthetic examples tend to preserve internal logic, requiring minor corrections.
RQ4 addressed the efficiency of data generation. With only 1,154 total LLM API calls—59 for generation and 1,095 for validation—we curated 618 high-quality samples, averaging just over two calls which costs only \$0.02 per accepted sample. This demonstrates that large-scale, diverse datasets can be created with minimal manual effort, adaptable to other structured domains with simulated task-specific errors.

Despite these strengths, limitations remain. The LLM-based reviewer agent, while scalable, diverges from human annotators (Cohen’s kappa 0.25), reflecting the challenge of modeling subjective plausibility. Reviewer judgments may misalign with human correctness, especially in ambiguous or multi-step reasoning cases. Additionally, narrow context windows due to token limits can omit globally relevant spreadsheet information like distant dependencies or multi-sheet references, limiting repair accuracy for complex scenarios. Finally, our dataset focuses on single-sheet, English-language spreadsheets, omitting collaborative, localized, and dynamic formula contexts common in practice.

These findings suggest promising future directions: incorporating human-in-the-loop validation to improve fidelity; refining reviewer agents via feedback fine-tuning or preference learning; enhancing context selection with structure-aware retrieval to prioritize semantically relevant spreadsheet regions; and extending the pipeline to multi-sheet, shared, and localized spreadsheets to broaden applicability.

Beyond benchmarking, \forepbench{} can serve as a data augmentation tool to improve formula understanding and repair models. By simulating common spreadsheet errors at scale, the pipeline offers a valuable resource for training and evaluating systems that assist users in real-world spreadsheet environments. Our work demonstrates that scalable, semi-automated generation of Excel formula repair data is feasible and effective. While synthetic data cannot fully replace human-curated examples, carefully filtered synthetic samples can enhance model robustness and enable systematic benchmarking. We hope \forepbench{} will support future research in end-user programming, intelligent assistants, and robust Excel formula repair.

% \subsection{Ethical Considerations}
% The data collection process involved selecting threads from the MrExcel forum. This is a publicly accessible platform, and our data collection adheres to the forum's usage policies, ensuring compliance with ethical standards and respect for user privacy.

% \begin{acks}
% This work was conducted with support from the Microsoft Excel team.
% \end{acks}

\bibliographystyle{ACM-Reference-Format}
\bibliography{sample-base}

\end{document}